\documentclass[aps,prc,twocolumn,amsmath,amssymb,floatfix,nofootinbib]{revtex4}
\usepackage{epsfig,colordvi}
\usepackage{rotating}
\usepackage{url}
\usepackage{graphicx}
\usepackage{dcolumn}
\usepackage{bm}
\usepackage{epsfig}
\usepackage{amsmath}
\usepackage{makeidx}
\usepackage[all,dark]{draftcopy}

\def\be{\begin{equation}}

\def\ee{\end{equation}}
\def\barr{\begin{array}}
\def\earr{\end{array}}

\def\nn8{\nonumber\\[2pt]}
\def\l{\left}
\def\r{\right}
\def\dis{\displaystyle}
\def\ed{\end{document}}

\begin{document}
\title{Event rates for the scattering of weakly interacting massive particles \\ from $^{23}$Na and $^{40}$Ar}

\author{R. Sahu$^{1}$\footnote{rankasahu@gmail.com}, 
V.K.B. Kota$^2$\footnote{vkbkota@prl.res.in}}

\affiliation{$^{1}$ National Institute of Science and Technology,
Palur Hills, Berhampur-761008, Odisha, India}

\affiliation{$^{2}$Physical Research Laboratory, Ahmedabad 380 009, India}

\date{\hfill \today}

\date{\hfill \today}
\begin{abstract}

Detection rates for the elastic and inelastic scattering of weakly
	interacting massive particles (WIMP) off 
	$^{23}$Na
	are calculated within the framework of Deformed Shell Model (DSM)
	based on Hartree-Fock states. First the spectroscopic properties
	like energy spectra and magnetic moments are calculated and compared
	with experiment. Following the good agreement for these, DSM wave functions are used for obtaining elastic and inelastic 
	spin structure functions, nuclear
	structure coefficients etc. for the WIMP-$^{23}$Na scattering.
	Then, the event rates are also calculated with a given set of
	supersymmetric parameters. In the same manner, using DSM wavefunctions, 
	nuclear structure coefficients and event rates for elastic scattering of WIMP from $^{40}$Ar are also obtained. These  results for event rates and also for annual modulation will be useful for the upcoming and future 
 WIMP detection	experiments involving detectors with  $^{23}$Na 
	and $^{40}$Ar.
	
\end{abstract}


\maketitle
\section{Introduction}

There is now universal agreement among the cosmologists, astronomers
and physicists that most of the mass of the universe is 
dark \cite{Jung-96,Hinshaw-2013,
Aghanim-2018}.
There are overwhelming evidences to believe that the dark matter
is mostly nonbaryonic. Also, data from the Cosmic
Background Explorer (COBE) \cite{Smoot-1992}  and Supernova Cosmology project
\cite{Gawser-1988} suggest that most of the dark matter is cold. 
The nonbaryonic cold dark matter is not yet observed in earth-bound 
experiments and hence its nature is still a mystery.  Axions are one of the candidates for dark matter but they are not yet observed \cite{Jung-96,admx-2018}. However,
the most promising nonbaryonic cold dark matter candidates are the Weakly 
Interacting Massive Particles (WIMP) which arise in super symmetric theories
of physics beyond the standard model. The most appealing WIMP candidate is
the Lightest Supersymmetric Particle (LSP) (lightest neutralino) which is expected to be stable and
interacts weakly with matter \cite{Jung-96, Verg-1996}.

There are many experimental efforts \cite{Freese-2013, Liu-2017, PRC-new, new-expts} to detect WIMP via their scattering
from the nuclei of the detector providing finger-prints regarding their existence. Some of these are Super CDMS SNOLAB project, XENON1T, PICO-60, EDELWEISS and so on; see for example \cite{new-expts, xenon1t,pico-2017,edelweiss}.
Nuclei $^{23}$Na, $^{40}$Ar, $^{71}$Ga, $^{73}$Ge, $^{75}$As, $^{127}$I, 
$^{133}$Cs and $^{133}$Xe are among the popular detector nuclei; see  
\cite{Divari-2000,new-expts,PRC-new} and references there in. Our focus in this paper is on $^{23}$Na and $^{40}$Ar. 
The Sodium Iodide (NaI) Advanced  Detector (NAID) array experiment is a direct
search experiment for WIMP operated by UK Dark Matter Collaboration 
in North Yorkshire \cite{NAIAD}; the NaI contains $^{23}$Na. 
Similarly, the DAMA/NaI and DAMA/LiBRA \cite{Bernabei-2018}
experiments investigated the presence of dark matter particles in the galactic
halo using the NaI(Tl) detector. In these experiments, the predicted annual modulation was not yet confirmed \cite{new-expts}. 
Other related experiments with NaI detectors are ANAIS \cite{ANAIS} 
and DM-Ice \cite{Ice}. Also, there are the important DARKSIDE-50 \cite{ref1} and DEAP-3600 \cite{ref2} experiments using lquid Argon (with $^{40}$Ar) as detector. 

Let us add that direct detection experiments are exposed
to various neutrino emissions. The
interaction of these neutrinos especially the astrophysical neutrinos 
with the material of the dark matter detectors known as the neutrino floor is a serious background source. 
Recently the coherent elastic scattering of neutrinos
off nuclei (CE$\nu$NS) has been observed at the Spallation Neutron Source
at the  Oak Ridge National Laboratory \cite{Akimov}
employing the technology used in the direct detection of dark matter searches. 
The impacts of the neutrino floor on the relevant experiments looking for cold
dark matter was investigated for example in \cite{nu-floor}.

There are many theoretical calculations which describe different aspects of 
direct detection of dark matter through the recoil of the nucleus in 
WIMP-nucleus scattering. For elastic scattering, we need to consider spin-spin
interaction coming from the axial current and also the more dominant scalar 
interaction. For inelastic part, scalar interaction practically
does not contribute. 
The scalar interaction can arise from squark exchange, Higgs exchange, 
the interaction of WIMPs with gluons etc.
Suhonen and his collaborators
have performed a series of truncated shell model calculations for this purpose
\cite{Suhonen-2004, Suhonen-2006, Suhonen-2009, Verg-2015, Suhonen-2016}. In these studies, for example they have calculated the
event rates for WIMP-nucleus elastic and inelastic scattering 
for $^{83}$Kr and $^{125}$Te \cite{Suhonen-2016} and also $^{127}$I, $^{129,131}$Xe 
and $^{133}$Cs \cite{Suhonen-2009}. In addition, recently
Vergados et al \cite{Verg-2018} examined the possibility of 
detecting electrons in the searches for light WIMP with a mass in the MeV region and found that the events of
0.5-2.5 per kg-y would be possible.
Few years back  full large-scale shell-model calculations are carried out in 
\cite{Divari-2000,Klos-2013} for WIMP
scattering off $^{129,131}$Xe, $^{127}$I, $^{73}$Ge, $^{19}$F, $^{23}$Na,
$^{27}$Al and $^{29}$Si nuclei. Finally, using large scale shell model \cite{ar40-1} and coupled cluster theory \cite{ar40-2} WIMP-nucleus and neutrino-nucleus scattering respectively, with $^{40}$Ar, are studied.  

In recent years, the deformed shell model (DSM), based on Hartree-Fock (HF) 
deformed intrinsic states with angular momentum projection and band mixing, 
has been established to be a good model to describe the properties of
nuclei in the mass range A=60-90 \cite{ks-book}. Among many applications, DSM is found to be quite  successful in describing
spectroscopic properties of medium heavy N=Z odd-odd nuclei with isospin projection
\cite{ga62-2015}, double beta decay
half-lives  \cite{SSK, dbd-2015} and  $\mu-e$ conversion in the field of the
nucleus \cite{mu-e-2003}. Going beyond these applications,
recently we have studied the event rates for WIMP with $^{73}$Ge as the
detector \cite{SK-2017}. In addition to the energy spectra
and magnetic moments, the model is used to calculate the spin structure 
functions, nuclear structure factors for the elastic and inelastic scattering.
Following this successful study, we have recently used DSM for calculating the neutrino-floor
due to coherent elastic
neutrino-nucleus scattering (CE$\nu$NS) \cite{nu-floor} for the candidate nuclei $^{73}$Ge, $^{71}$Ga, $^{75}$As and $^{127}$I. 
We found that the neutrino-floor 
contributions may lead to a distortion of the expected recoil spectrum limiting the sensitivity of the direct dark matter search experiments.  
In \cite{PRC-new}, DSM results for WIMP scattering from  $^{127}$I, $^{133}$Cs and $^{133}$Xe are described
in detail. To complete these studies that use DSM for the nuclear structure part, in the present paper we will present results for WIMP-
$^{23}$Na elastic and inelastic scattering and WIMP-$^{40}$Ar elastic scattering. Now we will give a preview.

Section II gives, for completeness and easy reading of the paper, 
a brief discussion of the formulation of WIMP-nucleus 
elastic and inelastic scattering and event rates. In Section III the DSM 
formulation is described with examples drawn from $^{75}$As spectroscopic 
results. In Section IV, spectroscopic results and also the
results for  elastic and inelastic scattering of WIMP from 
$^{23}$Na are presented. Similarly, WIMP-$^{40}$Ar elastic scattering results are presented in Section V. The results in Sections IV and V are the main results of this paper.
Finally, concluding remarks are drawn in Sect. VI.

\section{Event rates for WIMP-nucleus scattering}

WIMP flux on earth coming from the galactic halo is expected to be quite 
large, of the order $10^5$ per $cm^2$ per second. Even though the interaction
of WIMP with matter is weak, this flux is sufficiently large for the 
galactic WIMPs to deposit a measurable amount of energy in an appropriately 
sensitive detector apparatus when they scatter off nuclei. Most of the
experimental searches of WIMP is based on the 
direct detection through their interaction with nuclei in the detector.
The relevant  theory of WIMP-nucleus scattering is well known as available in the papers by Suhonen and his group and also in our earlier papers mentioned above
\cite{SK-2017, Suhonen-2016,Suhonen-2009,Suhonen-2006,Suhonen-2004}.
For completeness we give here a few important steps.
In the case of spin-spin interaction, the WIMP couples to the spin of the 
nucleus and in the case of scalar interaction, the WIMP couples to the mass of the nucleus.
In the expressions for the event rates, the super-symmetric part is  
separated from the nuclear part  so that the role played by the nuclear 
part becomes apparent. 

\begin{figure}
\includegraphics[width=6cm]{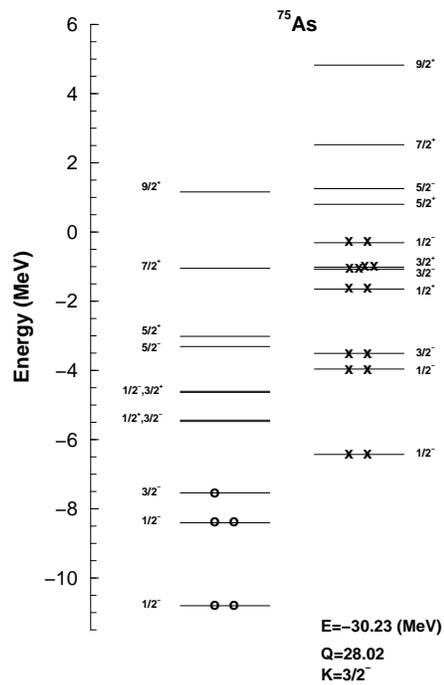}
        \caption{ HF single-particle spectra for $^{75}$As corresponding to
        lowest prolate configuration. In the figure, circles represent protons
        and crosses represent neutrons. The HF energy E in MeV, mass
        quadrupole moment Q in units of the square of the oscillator length
        parameter and the total azimuthal quantum number K are given
        in the figure.
}
\label{as75hf}
\end{figure}
\begin{figure}
        \includegraphics[width=6cm]{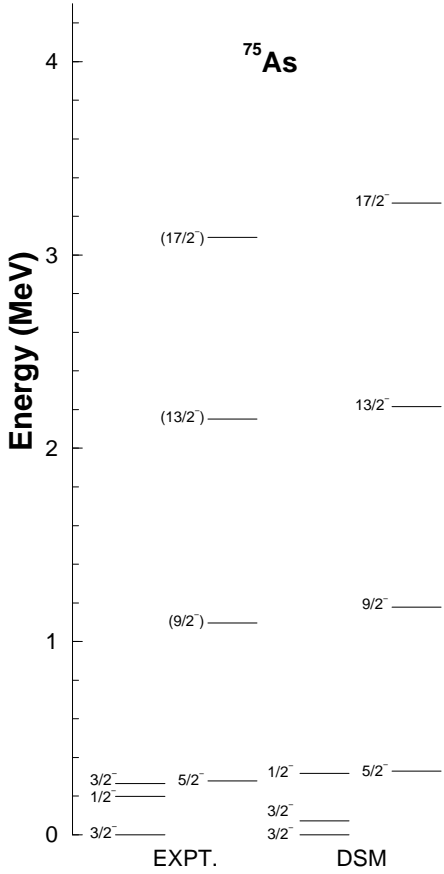}
\caption{ Comparison of DSM results with experimental data
        for $^{75}$As for collective bands with negative parity.
	The experimental values are taken from \cite{nndc}
}
\label{spect-as75}
\end{figure}

\subsection{Elastic scattering}

The differential event rate per unit detector mass for a WIMP with mass
$m_\chi$ can be written as \cite{Jung-96},
\be
dR = N_t\; \phi\; f \; \frac{d\sigma}{d\mid q \mid ^2} d^3 v\; 
d\mid q \mid^2
\label{eqn.1}
\ee
Here, $\phi$ which is equal to $\rho_0 v/m_\chi$
is the dark matter flux with $\rho_0$ being the  local WIMP density.
Similarly, $N_t$ stands for the number of target nuclei per unit mass and $f$ is the WIMP velocity distribution 
which is assumed to be Maxwell-Boltzmann type.
It takes into account the distribution of the WIMP velocity 
relative to the detector (or earth) and also the motion of the sun and earth. 
If we neglect  the
rotation of the earth about its own axis, then $v=\mid {\mathbf v}\mid$ is the 
relative velocity of WIMP with respect to the detector. Also, $q$ represents the
momentum  transfer to the nuclear target which is related to the dimensionless
variable  $u=q^2b^2/2$ with $b$ being the oscillator length parameter.  The
WIMP-nucleus differential cross section in the laboratory frame is given by
\cite{SK-2017, Suhonen-2016,Suhonen-2009,Suhonen-2006,Suhonen-2004}
\be
\frac{d\sigma (u,v)}{du} = \frac{1}{2}\, \sigma_0\,\left(\frac{1}{m_pb}
\right)^2 \,\frac{c^2}{v^2} \,\frac{d\sigma_{A}(u)}{du} \;;
\label{eqn.2}
\ee
with
%

\begin{equation}
\begin{aligned}
\dis\frac{d\sigma_{A}(u)}{du}  =  & (f_A^0)^2 F_{00}(u) +
2f_A^0 f_A^1  F_{01}(u) + (f_A^1)^2 F_{11}(u) \\   & +
\left[ Z \l(f_S^0 + f_S^1 \r) \right]^2 |F_Z(u)|^2 \\& + \left[ (A-Z) \l(f_S^0 - f_S^1 \r) \right]^2 |F_N(u)|^2 \\ & + 2 Z (A-Z) \left[(f_S^0)^2 - (f_S^1)^2 \right] |F_Z(u)||F_N(u)| \, .
\label{eqn.3}
\end{aligned}
\end{equation}
where $F_Z(u)$ and $F_N(u)$ denote the nuclear form factors for protons and neutrons respectively.
In Eq. (\ref{eqn.3}), the first three terms correspond to spin contribution
coming mainly from the axial current and the other three terms stand for the 
coherent part coming mainly from the scalar interaction. 
Here, $f_A^0$ and $f_A^1$ represent isoscalar and isovector parts of the axial
vector current and similarly  $f_S^0$ and $f_S^1$ represent isoscalar and
isovector parts of the scalar current. The nucleonic current parameters  $f_A^0$
and $f_A^1$ depend on the specific SUSY model employed.  However, $f_S^0$ and
$f_S^1$ depend,  beyond SUSY model, on the hadron model used to embed 
quarks and gluons into nucleons. 
The normalized spin structure functions $F_{\rho\rho'}(u)$ with
$\rho$, $\rho'$ = 0,1 are defined as
\be
\barr{l}
F_{\rho\rho'}(u) = \dis\sum_{\lambda,\kappa}\frac{\Omega_\rho^
{(\lambda,\kappa)}(u)\Omega_{\rho'}^{(\lambda,\kappa)}(u)}{\Omega_\rho
\Omega_{\rho'}}\;;\\
\\
\Omega_\rho^{(\lambda,\kappa)}(u) = \sqrt{\frac{4\pi}{2J_i + 1}} \\
                \times \langle J_f \| \dis\sum_{j=1}^A \left[Y_\lambda(\Omega_j)
\otimes \sigma(j)\right]_\kappa j_\lambda(\sqrt{u}\,r_j) 
\omega_\rho(j) \|J_i\rangle
\earr \label{eqn.6}
\ee 
In the above equation $\omega_0(j)=1$ and $\omega_1(j)=\tau(j)$; 
note that $\tau=+1$ for protons and $-1$ for neutrons  
and $j_\lambda$ is the spherical Bessel function. 
The static spin matrix elements are defined as $\Omega_\rho(0) =
\Omega_\rho^{(0,1)}(0)$. 
Now, the event rate can be written as 
\begin{equation}
\langle R \rangle = \int^1_{-1} d\xi  \int^{\psi_{max}}_{\psi_{min}} d\psi 
\int^{u_{max}}_{u_{min}}
G(\psi, \xi) \frac{d\sigma_{A}(u)}{du} du 
\label{eqn.9}
\end{equation}
In the above, $G(\psi, \xi)$ is given by
\begin{equation}
G(\psi, \xi) = \frac{\rho_0}{m_\chi} \frac{\sigma_0}{Am_p} \left(\frac{1}
{m_pb}\right)^2 \frac{c^2}{\sqrt{\pi}v_0} \psi e^{-\lambda^2} e^{-\psi^2}
e^{-2\lambda\psi\xi}
\label{eqn.10}
\end{equation}
Here, $\psi=v/v_0$, $\lambda=v_E/v_0$, $\xi=cos(\theta)$.
Parameters used in the calculation are the following: the WIMP density
$\rho_0 = 0.3 \;Gev/{cm^3}$, $\sigma_0 = 0.77 \times 10^{-38} cm^2$, mass
of proton $m_p = 1.67 \times 10^{-27}$ kg. The
velocity of the sun with respect to the galactic centre is taken to be
$v_0 =220$ Km/s and the velocity of the earth relative to the sun is taken as
$v_1=30$ Km/s. The velocity of the earth with respect to the galactic
centre $v_E$ is given by
$v_E = \sqrt{v_0^2 + v_1^2 + 2v_0v_1 \sin(\gamma) cos(\alpha)}$
where $\alpha$ is the modulation angle which stands for the phase of the earth on  its
orbit around the sun and $\gamma$ is the angle between the normal to the
elliptic and the galactic equator which is taken to be $\simeq 29.8^\circ$. Using the notations,
$X(1)=  F_{00}(u)$,
$X(2)= F_{01}(u)$,
$X(3)=   F_{11}(u)$,
$X(4) = |F_Z(u)|^2$,
$X(5) = |F_N(u)|^2$,
$X(6) = |F_Z(u)||F_N(u)|$
the event rate per unit mass of the detector is given by 
\begin{equation}
\begin{aligned}
\langle R \rangle_\text{el}  =&  (f^0_1)^2 D_1 + 2 f^0_Af^1_A D_2 + (f^1_A)^2 D_3  \\ 
& + \left[Z \l(f_S^0 + f_S^1 \r) \right]^2 D_4 \\ & +  \left[(A-Z) \l(f_S^0 - f_S^1 \r) \right]^2 D_5 \\ & + 2 Z (A-Z) \l[(f_S^0)^2 - (f_S^1)^2 \r] D_6 \, ,
\end{aligned}
\label{eqn.12}
\end{equation}
where $D_i$  being the three dimensional integrations of Eq.(\ref{eqn.9}), defined as
\be
D_i = \int^1_{-1} d\xi  \int^{\psi_{max}}_{\psi_{min}} d\psi 
\int^{u_{max}}_{u_{min}}
G(\psi, \xi) X(i) du 
\label{eqn.9a}
\ee
The lower and upper limits of integrations given in Eq.(\ref{eqn.9}) and 
(\ref{eqn.9a}) have been worked out by Pirinen et al \cite{Suhonen-2016} and 
they are 
\be
\psi_{min} = \frac{c}{v_0} \left(\frac{Am_pQ_{thr}}{2\mu^2_r}\right )^{1/2}
\label{eqn.13}
\ee
\be
\psi_{max} = -\lambda\xi + \sqrt{\lambda^2\xi^2+\frac{v_{esc}^2}{v_0^2} -1
- \frac{v^2_1}{v^2_0}-\frac{2 v_1}{v_0} sin(\gamma) cos(\alpha)}
\label{eqn.14}
\ee
With the escape velocity $v_{esc}$ from our galaxy to be 625 km/s, the
value of $v_{esc}^2/v_0^2 -1- v^2_1/v^2_0$ appearing in 
Eq. (\ref{eqn.14}) is $7.0525$. Similarly, the value
of $(2 v_1/v_0) sin(\gamma)$ is $0.135$. The values of $u_{min}$ and
$u_{max}$ are $Am_p Q_{thr}b^2$ and $2(\psi\mu_rbv_0/c)^2$, respectively.
Here, $Q_{thr}$ is the detector threshold energy and $\mu_r$
is the reduced mass of the WIMP-nucleus system.

\subsection{Inelastic scattering}

In the inelastic scattering the entrance channel and exit channel are different.
The inelastic scattering cross section due to scalar current is considerably 
smaller than the elastic case and hence it is neglected.
Hence, we focus on spin dependent scattering.  
The inelastic event rate per unit mass of the detector can be written as
\be
\langle R \rangle_{in} = (f^0_1)^2 E_1 + 2 f^0_Af^1_A E_2 + (f^1_A)^2 E_3
\label{eqn.19}
\ee
where $E_1$, $E_2$ and $E_3$ are the three dimensional integrations
\be
E_i = \int^1_{-1} d\xi  \int^{\psi_{max}}_{\psi_{min}} d\psi 
\int^{u_{max}}_{u_{min}}
G(\psi, \xi) X(i) du\;. 
\label{eqn.20}
\ee
The limits of integration for $E_1$, $E_2$ and $E_3$ are \cite{Suhonen-2016, Suhonen-2009} 
\be
u_{min(max)} = \frac{1}{2}b^2\mu_r^2\frac{v^2_0}{c^2}\psi^2
\left[ 1 \mp \sqrt{1-\frac{\Gamma}{\psi^2}} \right ]^2
\label{eqn.21}
\ee
where 
\be
\Gamma = \frac{2 E^*}{\mu_rc^2} \frac{c^2}{v_0^2}
\label{23}
\ee
with $E^*$ being the energy of the excited state. $\psi_{max}$ is same as in the
elastic case and the lower limit  $\psi_{min} = \sqrt{\Gamma}$. The 
parameters like $\rho_0$, $\sigma_0$ etc. have the same values as in the 
elastic case.

\section{Deformed shell model}

The nucleonic  current part has
been separated from nuclear part 
in the expression for the event rates for elastic and inelastic scattering 
given by Eqs.  (\ref{eqn.12}) and (\ref{eqn.19}) respectively with $X(i)$ giving the nuclear structure part. However,
the $D_i$'s and $E_i$'s depend 
not only on the nuclear structure
part but also on the kinematics and assumptions on the WIMP velocity. The
evaluation of $X(i)$  depends on  spin structure  functions  and the form factors.
We have used DSM for the evaluation of these quantities. Here, for a given
nucleus, starting with a model space consisting of a given set of single
particle (sp) orbitals and effective two-body Hamiltonian (TBME + spe), the
lowest energy intrinsic states are obtained by solving the Hartree-Fock (HF)
single particle equation self-consistently. We assume axial symmetry. 
For example, Fig. \ref{as75hf} shows the HF single particle spectrum 
for $^{75}$As corresponding to the lowest prolate intrinsic state. 
Used here are the spherical sp orbits  $1p_{3/2}$, $0f_{5/2}$,
$1p_{1/2}$, and $0g_{9/2}$ with energies 
0.0, 0.78, 1.08, and 3.20 MeV, respectively, while the assumed effective 
interaction is the modified Kuo interaction \cite{Kuo}. Excited
intrinsic configurations are obtained by making particle-hole excitations over
the lowest intrinsic state.  These intrinsic states $\chi_K(\eta)$ do not have
definite angular momenta.  Hence, states of good  angular momentum are
projected from
an intrinsic state $\chi_K(\eta)$ and they can be written as,
\begin{equation}
\psi^J_{MK}(\eta) = \frac{2J+1}{8\pi^2\sqrt{N_{JK}}}\int d\Omega D^{J^*}_{MK}(\Omega)R(\Omega)| \chi_K(\eta) \rangle 
\label{eqn.24}
\end{equation}
where $N_{JK}$ is the normalization constant.
In Eq. (\ref{eqn.24}), $\Omega$ represents the Euler angles ($\alpha$, $\beta$,
$\gamma$) and $R(\Omega)$ which is equal to exp($-i\alpha J_z$)exp($-i\beta
J_y$)exp( $-i\gamma J_z$) represents the general rotation operator.  The good
angular momentum states projected from different intrinsic states are not in
general orthogonal to each other.  Hence they are orthonormalized and then 
band mixing calculations are performed. This gives the energy spectrum and the eigenfunctions. Fig. \ref{spect-as75} shows the calculated energy spectrum for $^{75}$As as an example.  In the DSM band mixing calculations used are six intrinsic states \cite{nu-floor}. Let us add that the eigenfunctions are of 
the form
\be
\vert\Phi^J_M(\eta) \rangle \, =\,\sum_{K,\alpha} S^J_{K \eta}(\alpha)\vert 
\psi^J_{M K}(\alpha)\rangle \;.
\label{phijm}
\ee
The nuclear matrix elements occurring in the calculation of magnetic moments,
elastic and inelastic spin structure functions etc. are evaluated using 
the wave function 
$\Phi^J_M(\eta)$. For example the calculated magnetic moments for the
$3/2_1$, $3/2_2$ and $5/2_1$ states are (in $nm$ units) 1.422, 1.613 and 0.312 compared to experimental values \cite{nndc} 1.439, 0.98 and 0.98 respectively. The calculated values are obtained using bare gyromagnetic ratios and the results
will be better for the excited states if we take  $g_\ell^p=0.5$,
$g^n_{\ell}=0.7$, $g_s^p=4$ and $g_s^n=-3$.  The neutron spin part is small
and hence donot appreciably contribute to the magnetic moments of the above
three states. Use of effective $g$-factors are advocated in \cite{effect-g}.  

\begin{figure}
\includegraphics[width=6cm]{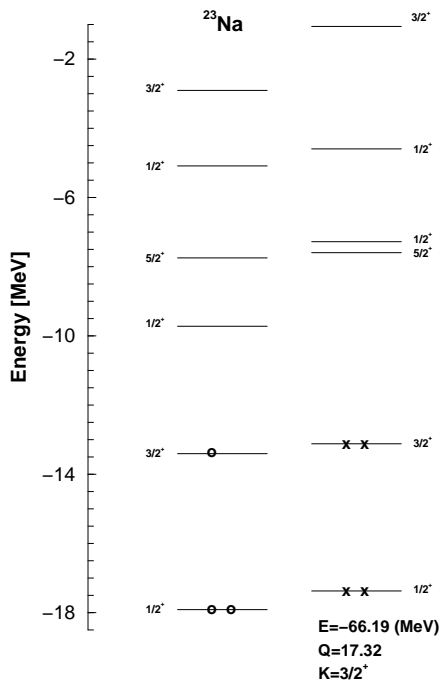}
        \caption{ HF single-particle spectra for $^{23}$Na corresponding to
        lowest configuration. In the figure, circles represent protons
        and crosses represent neutrons. The HF energy E in MeV, mass
        quadrupole moment Q in units of the square of the oscillator length
        parameter and the total azimuthal quantum number K are given
        in the figure.
}
\label{na23hf}
\end{figure}
\begin{figure}
        \includegraphics[width=4cm]{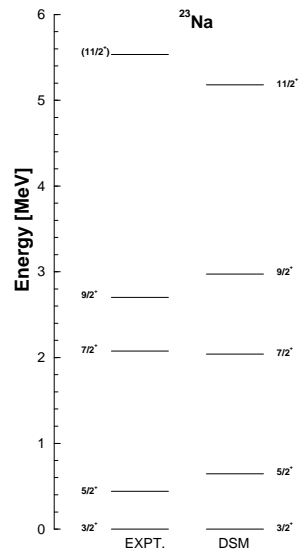}
\caption{ Comparison of deformed shell model results with experimental data
        for $^{23}$Na for yrast band which is of positive parity. 
	The experimental values are taken from \cite{nndc}
}
\label{spect_na23}
\end{figure}

\section{Results for WIMP-$^{23}$Na scattering}

The nuclear structure plays an important role in studying the event rates
in WIMP-nucleus scattering. Hence, we first calculate the energy spectra and
magnetic moments within our DSM model for $^{23}$Na. Agreement with experimental data will 
provide information regarding the goodness of the wave functions 
used.  This in turn will give us confidence regarding the reliability of 
our predictions on event rates. These spectroscopic results are presented in Section IV-A. Let us add that in Sections IV-B and C
the value of the oscillator length parameter $b$ is needed and it is taken to be 
1.573 fm for $^{23}$Na. In our earlier
work in the calculation of transition matrix  elements for $\mu - e$ conversion
in $^{72}$Ge \cite{mu-e-2003}, we had taken the  value of this length parameter
as 1.90 fm. Assuming $A^{1/6}$ dependence, the above values of the oscillator
parameter is chosen for $^{23}$Na.  

\subsection{Spectroscopic results}

In the $^{23}$Na calculations, $^{16}$O is taken as the
inert core with  the spherical single particle orbitals $0d_{5/2}$, $1s_{1/2}$ and $0d_{3/2}$ generating the basis space. "USD" interaction of Wildenthal with sp energies $-3.9478$, $-3.1635$ and $1.6466$ 
 MeV has been used in the calculation \cite{sd-shell}.
This effective interaction is known to be quite successful in describing most of the important 
spectroscopic features
of nuclei in the $1s0d$-shell region \cite{sd-shell}.  For this nucleus, the calculated lowest HF single 
particle spectrum of prolate shape is shown in Fig. \ref{na23hf}. The odd proton
is in the $k= 3/2^+$ deformed single particle orbit. The excited
configurations are obtained by particle-hole excitations over this lowest
configuration. We have considered 
a total of five intrinsic configurations. As described above, angular momentum
states are projected from each of these intrinsic configurations and then a 
band mixing calculation is performed. The band mixed wave functions 
$S^J_{K \eta}$ defined in Eq. (\ref{phijm}) are
used to calculate the energy levels, magnetic moments and other properties
of this nucleus. 

The calculated levels are classified into collective bands
on the basis of the E2 transition probabilities between them. The results for 
lowest positive parity band for $^{23}$Na are shown in Fig. \ref{spect_na23}.
The experimental data are from Ref. \cite{nndc}. For this nucleus, the
ground state is $3/2^+$ which is reproduced in our calculation. A positive
parity band built on $3/2^+$ has been identified for this nucleus. This
band is quite well reproduced by the DSM calculation. 
An analysis of the wave
functions shows that this band mainly originates from the lowest HF intrinsic
configuration shown in Fig. \ref{na23hf}.  
However, there are admixtures
from the good angular momentum states coming from other intrinsic 
configurations. 
The wavefunction coming from the lowest HF intrinsic configuration
slightly increases in value with increased angular momentum. This shows
that the collectivity of this band does not change appreciably at higher
angular momentum.
Since we are considering WIMP-nucleus scattering
from ground state and low lying positive parity states, the negative parity
bands are not important for the present purpose.

In the calculation of the event rates, spin plays an important role. Hence, magnetic moment of various low-lying levels in $^{23}$Na
are calculated. The result for the ground state of the lowest $K=3/2^+$
band is 2.393 
nm and the corresponding
available experimental data value is 2.218 nm. The contribution of protons
and neutrons to the orbital parts are 0.957 and 0.262 and to the spin parts
are 0.267 and 0.014, respectively.
This decomposition gives better physical insight. The calculated value of 
magnetic moment for the ground state agrees quite well with experimental data \cite{nndc}. Let us add that
there are no experimental data for the magnetic moments of the excited states.
An approach with state-dependent gyromagnetic moments, as advocated for 
example in \cite{effect-g} reproduces better the experimental magnetic 
moments. The DSM spectroscopic results are also in good agreement with the 
full shell model calculations reported in \cite{Divari-2000,Klos-2013}. 

\begin{figure}
\includegraphics[width=4.5cm]{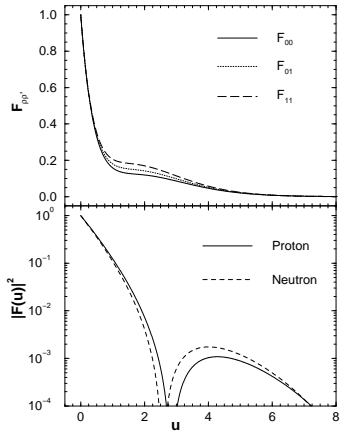}
\caption{Spin structure functions  and squared proton
and neutron form factors for $^{23}$Na for the ground state.}
\label{na23-el-ssf}
\end{figure}
\begin{figure}
\includegraphics[width=8cm]{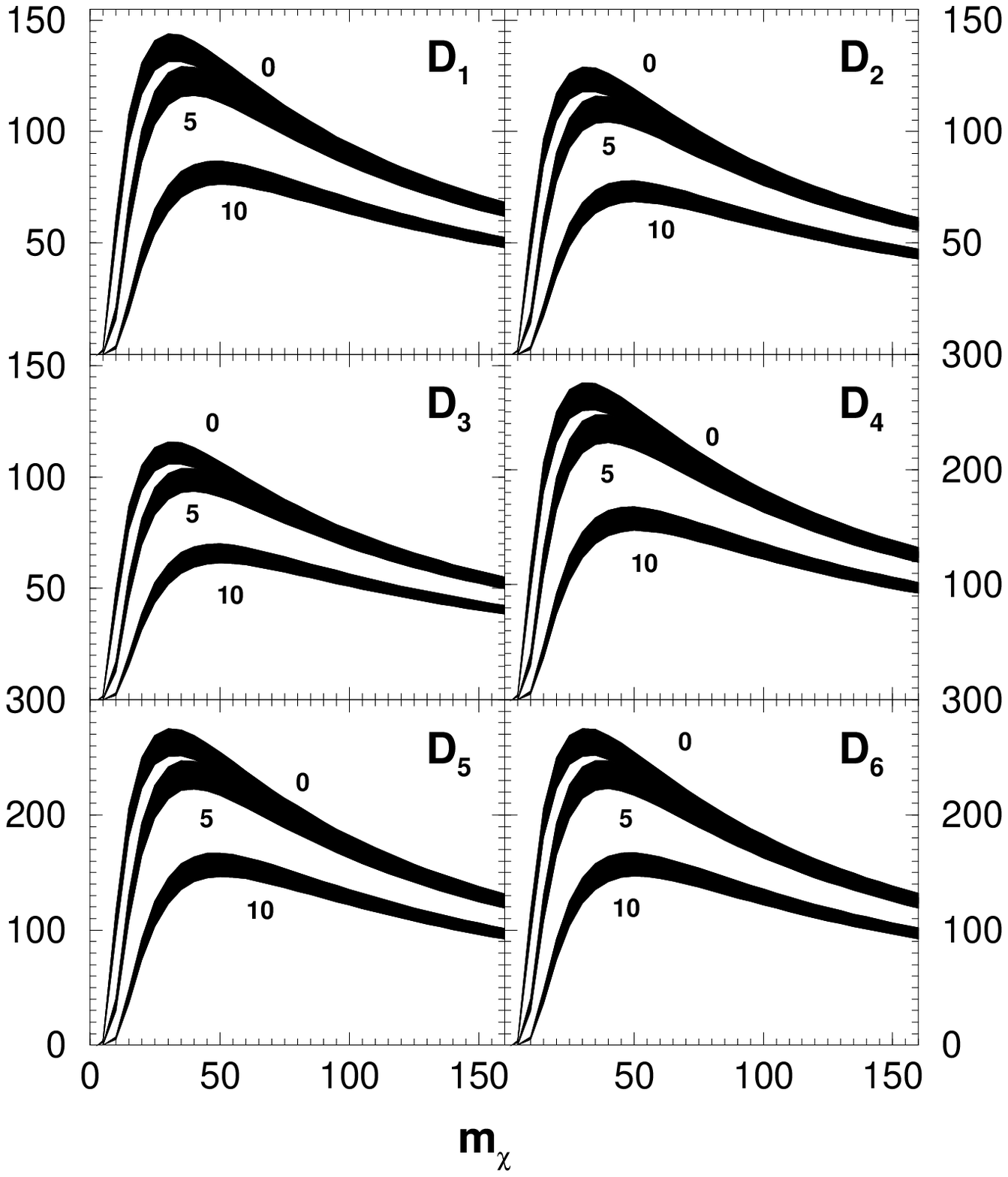}
\caption{Nuclear structure coefficients plotted as a function
        of the WIMP mass in GeV for $^{23}$Na.
        The graphs are plotted for three values of the
detector threshold $Q_{thr}$ namely $Q_{thr}=0,5,10$ keV. The
thickness of the graphs
for each value of $Q_{thr}$ represents the annual modulation.
}
\label{dn_el_na23}
\end{figure}
\begin{figure}
\includegraphics[width=6.5cm]{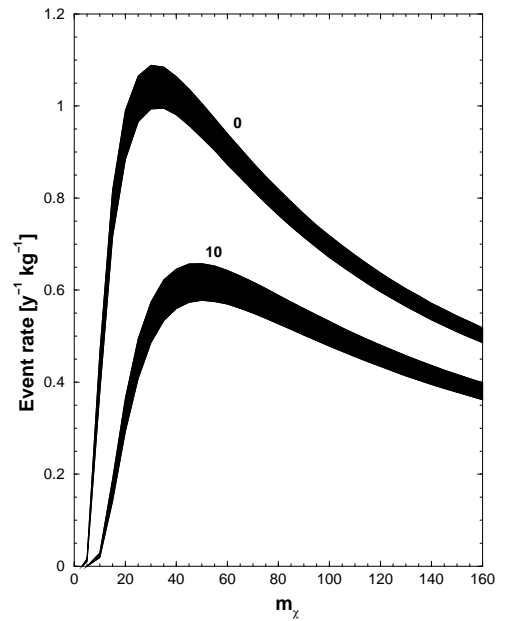}
\caption{The event rates in units of $yr^{-1} kg^{-1}$ as a function of
	dark matter mass in GeV for $^{23}$Na at detector threshold
	$Q_{th}=0, 10 \; keV$
	The thickness of the curves
 represent the annual modulation.
}
\label{rate_na23}
\end{figure}
 
\subsection{Results for elastic scattering}

The DSM wave functions given by Eq. (\ref{phijm}) are used to calculate
the normalized spin structure 
functions given in Eq. (\ref{eqn.6}) and also the squared nuclear form
factors for these nuclei. Their values are plotted in Figs. 
\ref{na23-el-ssf}
as a function of $u$.  The static spin matrix elements
$\Omega_0$ and $\Omega_1$ for the ground state of $^{23}$Na have values
0.727 and 0.652 respectively. They compare well with other theoretical calculations for $^{23}$Na given in \cite{Divari-2000,Klos-2013}.
An analysis of the normalized spin structure functions for $^{23}$Na
in Fig. \ref{na23-el-ssf} shows that the values of $F_{00}$, $F_{01}$ and 
$F_{11}$ differ  between u=0.4-3. Out side this region they are almost
degenerate. The form factors
for proton and neutron in $^{23}$Na are almost identical up to $u=2$. Afterwards
they differ and beyond u=2.6
the neutron form factor becomes larger than proton form factor.

The nuclear structure dependent coefficients given in Eq. (\ref{eqn.9a}) are plotted in Fig. \ref{dn_el_na23} for $^{23}$Na, as a function of the WIMP mass for different values of the detector
threshold. Since $\Omega_0$ and $\Omega_1$
are of same sign, $D_i$s are all positive. 
The peaks of the nuclear structure coefficients occur at around $m_\chi\sim$
30 GeV at zero threshold energy. The peaks shift towards higher values of 
$m_\chi$ as we go to larger threshold energy.
The thickness of the graphs represents annual modulation. Annual modulation has
largest value near the peaks of the graphs. 
Annual modulation provides strong evince regarding the observation of dark
matter since the back ground does not exhibit such modulation; see \cite{new-expts} for a recent review on annual modulation measurements. As seen from Figs. 6 and 7,  
$^{23}$Na shows larger modulation compared to heavier nuclei like $^{127}$I,
$^{133}$Cs and $^{133}$Xe \cite{PRC-new}.

The event detection rates for these nuclei have been calculated at a particular
WIMP mass by reading out the corresponding values of $D_i$s from the Fig. \ref{dn_el_na23} and then evaluating Eq. (\ref{eqn.12}) for
a given set of supersymmetric parameters. The event
detection rates for different values of $m_\chi$ have been 
calculated using the nucleonic current  parameters $f^0_A=3.55e-2$,
$f^1_A=5.31e-2$, $f^0_S=8.02e-4$ and $f^1_S=-0.15\times f^0_S$. 
These results are shown in Fig. (\ref{rate_na23}) for detector threshold energy 
$Q_{th}=$ 0, 10 keV for $^{23}$Na. 
For $^{23}$Na, the peak occurs at $m_\chi \simeq 30$ GeV. The event rate decreases 
at higher detector threshold energy but the peak shifts to the higher
values of $m_\chi$ occurring at $\sim$50 GeV. 

\begin{figure}
\includegraphics[width=4cm]{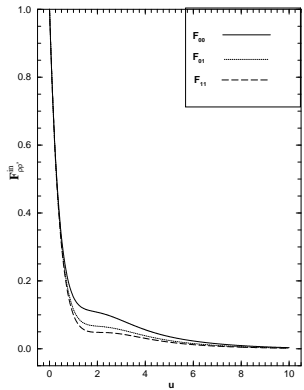} 
\caption{Spin structure function in the inelastic channel
	$5/2^+ \rightarrow 3/2^+$ 
	for $^{23}$Na.
\label{ssf_in_na23}
}
\end{figure}

\begin{figure}
\includegraphics[width=5cm]{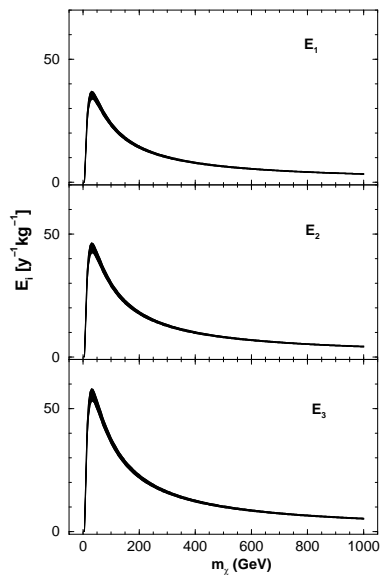} 
\caption{Nuclear structure coefficients $E_n$ in the inelastic channel
	$5/2^+ \rightarrow 3/2^+$ 
	for $^{23}$Na.
The thickness of the graphs represents annual modulation.
}
\label{ee_na23}
\end{figure}

\subsection{Results for inelastic scattering}

$^{23}$Na has $5/2^+$ excited state at 440 KeV above the ground state $3/2^+$. 
Therefore, we consider inelastic scattering from the
ground state for this nucleus to the $5/2^+$ state. The static spin matrix elements for the 
inelastic scattering to the $J=5/2^+$ are $\Omega_0 = -0.368$, 
$\Omega_1 = -0.462$. These values are
of the same order of magnitude as for the elastic scattering case. Again
$\Omega_0$ and $\Omega_1$ are of same sign.
The inelastic spin structure functions are given in Fig. 
(\ref{ssf_in_na23}). In the figures, $F_{00}$, $F_{01}$ and $F_{11}$ are shown.
The spin structure functions
almost vanish above u=4. With the value of $u$ lying between 1 to 4, the spin
structure functions differ from each other. The nuclear structure
coefficients $E_n$ are shown in Fig. \ref{ee_na23} for this nucleus. 
The inelastic nuclear structure coefficients do not depend on the detector
threshold energy. Hence the event rate can be calculated by reading the
values of $E_i$ from the graph and using the nucleonic current parameters.
The modulation for the inelastic scattering case is much smaller than the
elastic case.
\begin{figure}
\includegraphics[width=8.5cm]{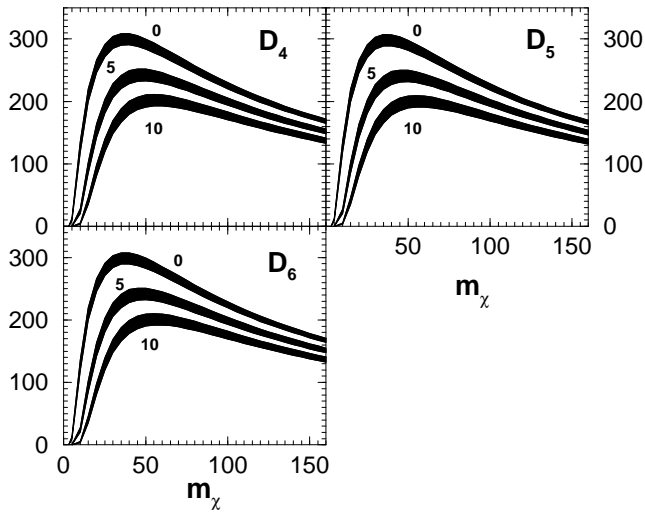}
\caption{Nuclear structure coefficients plotted as a function
        of the WIMP mass $m_\chi$ in GeV for $^{40}$Ar.
        The graphs are plotted for three values of the
detector threshold $Q_{thr}$ namely $Q_{thr}=0,5,10$ keV. The
thickness of the graphs
for each value of $Q_{thr}$ represents the annual modulation.
}
\label{dn_el_ar40}
\end{figure}
\begin{figure}
\includegraphics[width=6.5cm]{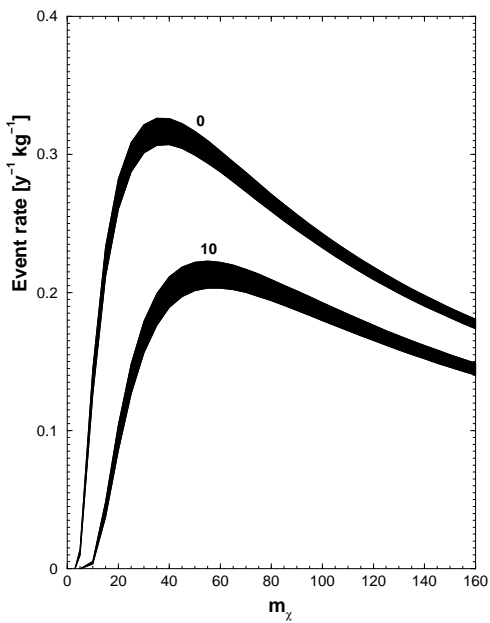}
\caption{The event rates in units of $yr^{-1} kg^{-1}$ as a function of
        dark matter mass $m_\chi$ in GeV for $^{40}$Ar at detector threshold
        $Q_{th}=0, 10 \; keV$
        The thickness of the curves
 represent the annual modulation.
}
\label{rate_ar40}
\end{figure}
\section{Results for WIMP-$^{40}$Ar Elastic Scattering}
The event rates for WIMP-$^{40}$Ar elastic scattering are calculated using
the nuclear wave functions generated through our DSM calculation. In our
calculation, the active spherical single particles orbitals are taken as
$0d_{5/2}$, $0d_{3/2}$, $1s_{1/2}$, $0f_{7/2}$, $0f_{5/2}$, $1p_{3/2}$ and
$1p_{1/2}$ with $^{16}$O as the inert core. An effective interaction named
$sdpf-u$ and developed by Nowacki and Poves \cite{Nowacki-2009} with
single particle energies $-3.699$, $1.895$, $-2.915$, $6.220$, $11.450$, 
$6.314$ and $6.479$ MeV, respectively
for the above seven orbitals has been used. As discussed
earlier, we first generate the lowest HF intrinsic state by solving the
axially symmetric HF equation self-consistently. Then, we generate the excited
configurations by particle-hole excitations. We have
considered a total of 9 intrinsic states. Good angular momentum states are
projected from each of these intrinsic states and then a band mixing 
calculation is performed. The band mixed wave functions defined in Eq.
(\ref{phijm}) are used in calculating the elastic event rates and
nuclear structure coefficients for the ground state of this nucleus. Note that the ground state is a $0^+$ state as $^{40}$Ar is a even-even nucleus and inelastic scattering from ground needs excited $1^+$ state. However, the $1^+$ states lie very high in energy and hence only elastic scattering of WIMP from $^{40}$Ar is important.
The oscillator length parameter $b$ for this nucleus is taken to be 1.725 $fm$.
We have presented the nuclear structure dependent coefficients defined
in Eq. (\ref{eqn.9a}) in Fig. \ref{dn_el_ar40} for this nucleus
as a function of the WIMP mass for different values of the detector threshold.
Since $^{40}$Ar is a even-even nucleus, there is no spin contribution from the
ground state. Hence, we have only $D_4$, $D_5$ and $D_6$ corresponding to the 
proton, neutron and proton-neutron form factors as defined in Eq. 
(\ref{eqn.12}). Their values are slightly larger compared to the corresponding 
quantities in $^{23}$Na but the modulation is smaller. The peaks occur at around
$m_\chi = 35$ GeV. However for larger values of $Q_{thr}$, the peaks shift
towards the larger $m_\chi$. 
The event rates for WIMP-$^{40}$Ar scattering is plotted as
a function of the dark matter mass in Fig. (\ref{rate_ar40}) for $Q_{thr}$ = 0 
and 10 keV. The event rates are calculated using the same supersymmetric
parameters as in $^{23}$Na. The values are smaller than in $^{23}$Na. This
is because $^{40}$Ar is a even-even nucleus and hence there is no spin 
contribution to the event rates in the ground state. At $Q_{thr}$ = 0, the
peak occurs at 35 GeV. For $Q_{thr}$ = 10 keV, the peak shifts to 45 GeV.

\section{Conclusion}

Deformed  shell model is used to calculate first the event rates for
the elastic and inelastic scattering of WIMP from $^{23}$Na.
Spectroscopic properties of this nucleus are calculated within DSM
to check the suitability of the model. We have also calculated  magnetic moments
for the lowest level in this nucleus since spin plays an important role 
in the calculation of
detection rates. Before $^{23}$Na analysis, we have compared the DSM results also for $^{75}$As for further confirmation of the goodness of DSM for spectroscopy. After ensuring the good agreement
with experiment, we calculated the spin structure functions, form factors,
nuclear structure coefficients and the event rates for WIMP-$^{23}$Na elastic and inelastic scattering. In addition, event rates for elastic scattering of WIMP from $^{40}$Ar are also presented. Results in Figs. 7 and 11 for event rates and in Figs. 6,7 and 9-11 for the annual modulation should be useful for the upcoming and future experiments detecting WIMP involving detectors with  $^{23}$Na and $^{40}$Ar.
Let us add that the present study using DSM for the nuclear structure part is in addition to the results presented for WIMP scattering from $^{73}$Ge in \cite{SK-2017} and from $^{127}$I, $^{133}$Cs and $^{133}$Xe in \cite{PRC-new}. Finally, we hope that
these and those obtained using other theoretical models for nuclear structure may guide the experimentalists to
unravel the fundamental mysteries of dark matter particles.

\acknowledgements

Thanks are due to Prof. T.S. Kosmas for his interest in this work.
R. Sahu is thankful to SERB of Department of
Science and Technology (Government of India) for financial support.

\end{document}